\documentclass[conference]{IEEEtran}
\IEEEoverridecommandlockouts
\usepackage{cite}
\usepackage{amsmath,amssymb,amsfonts}
\usepackage{algorithmic}
\usepackage{graphicx}
\usepackage{textcomp}
\usepackage{xcolor}
\usepackage{booktabs}
\usepackage{breakurl} 
\usepackage{array} 
\usepackage{tabularx}
\usepackage{makecell} 

\usepackage{amsmath}
\usepackage{subfigure}
\usepackage{amsfonts}
\usepackage{amsthm}
\usepackage{bm}
\def\eqref#1{equation~\ref{#1}}
\def\1{\bm{1}}

\def\vb{{\bm{b}}}

\def\vf{{\bm{f}}}

\def\vs{{\bm{s}}}

\def\vx{{\bm{x}}}

\def\mA{{\bm{A}}}

\def\mD{{\bm{D}}}

\def\mF{{\bm{F}}}

\def\mH{{\bm{H}}}
\def\mI{{\bm{I}}}

\def\mP{{\bm{P}}}

\def\mS{{\bm{S}}}

\def\mW{{\bm{W}}}
\def\mX{{\bm{X}}}

\DeclareMathAlphabet{\mathsfit}{\encodingdefault}{\sfdefault}{m}{sl}
\SetMathAlphabet{\mathsfit}{bold}{\encodingdefault}{\sfdefault}{bx}{n}
\newcommand{\tens}[1]{\bm{\mathsfit{#1}}}

\def\tP{{\tens{P}}}

\def\BibTeX{{\rm B\kern-.05em{\sc i\kern-.025em b}\kern-.08em
    T\kern-.1667em\lower.7ex\hbox{E}\kern-.125emX}}
\begin{document}

\title{Deep Learning-based Binary Analysis for Vulnerability Detection in x86-64 Machine Code}

\author{\IEEEauthorblockN{Mitchell Petingola}
\IEEEauthorblockA{\textit{Department of Computer Science and Mathematics} \\
\textit{Algoma University}\\
Sault Ste Marie, Canada \\
mpetingola@algomau.ca}}

\maketitle

\maketitle

\begin{abstract}
While much of the current research in deep learning-based vulnerability detection relies on disassembled binaries, this paper explores the feasibility of extracting features directly from raw x86-64 machine code. Although assembly language is more interpretable for humans, it requires more complex models to capture token-level context. In contrast, machine code may enable more efficient, lightweight models and preserve all information that might be lost in disassembly. This paper approaches the task of vulnerability detection through an exploratory study on two specific deep learning model architectures and aims to systematically evaluate their performance across three vulnerability types. The results demonstrate that graph-based models consistently outperform sequential models, emphasizing the importance of control flow relationships, and that machine code contains sufficient information for effective vulnerability discovery. 
\end{abstract}


\section{Introduction}

Software security is a critical area of concern due to the growing complexity and frequency of cyberattacks exploiting vulnerabilities in code. Among the vulnerabilities commonly present in software systems, memory corruption vulnerabilities such as mishandling pointers and buffer overflows remain common targets for exploitation \cite{saitoSurveyPreventionMitigation2016}.

Traditional approaches to vulnerability detection primarily include manual code reviews, static analysis tools, and dynamic testing methodologies. Manual code reviews conducted by expert analysts have long been considered the most reliable method, using human intuition and expertise to uncover deeply hidden vulnerabilities, but are inherently labor-intensive, costly, and impractical to scale across large codebases. Heuristic-based static analysis tools offer greater scalability but are known to result in several false positives \cite{kangDetectingFalseAlarms2022, desilvaComparativeAnalysisStatic2023}.

Deep learning techniques have demonstrated remarkable success across various domains, including natural language processing, computer vision, and speech recognition, motivating their application in software analysis. Contemporary deep learning approaches for vulnerability detection have focused on analyzing disassembled binaries or intermediate code representations obtained from compiled binaries, which can provide a human-readable format beneficial for manual inspection and interpretation. However, analyzing assembly language often requires more complex natural language processing (NLP)-like techniques to capture complex semantic and contextual relationships, which, in turn, increases computational and data requirements. Additionally, the disassembly process itself could potentially introduce inaccuracies and information loss due to heuristic assumptions or inherent limitations of disassembly tools \cite{liGenerationDisassemblyGround2020, 8812038}. 

Analyzing raw machine code directly offers a compelling approach by circumventing any potential disassembly-induced inaccuracies and preserving complete information. Machine code represents the precise sequence of executable instructions as interpreted by hardware, providing an accurate and unambiguous representation of software behavior. Additionally, machine code's structural simplicity and consistency make it particularly suitable for constructing lighter and more computationally efficient deep learning models. Despite these advantages, the direct application of deep learning techniques to machine code remains relatively unexplored, and many questions remain unanswered regarding the effectiveness and practicality of machine code-based vulnerability detection compared to traditional assembly-based approaches.

The primary objective of this paper is to explore the feasibility of applying deep learning techniques directly to machine code for vulnerability detection. The research investigates whether deep learning models trained specifically on x86-64 machine code can achieve comparable or superior accuracy relative to previous approaches that use assembly language. This paper also explores how different model architectures and hyperparameters affect performance across specific vulnerability types, including mishandled pointers, array bound violations, and integer overflows, aiming to determine which vulnerability types benefit most from machine code-level analysis. As this approach is relatively unexplored, this paper also aims to identify clear directions for future research and tool development, potentially leading to more robust and efficient solutions.

The following section provides context by reviewing foundational concepts and examines existing research in deep learning-based binary analysis.

\section{Related Work}

\subsection{Memory Corruption Vulnerabilities}

Memory corruption vulnerabilities occur when a program unintentionally modifies the contents of memory due to programming errors \cite{saitoSurveyPreventionMitigation2016}. These errors can compromise the integrity of a program’s memory space, leading to undefined behavior or crashes. Attackers can exploit these types of vulnerabilities to manipulate program execution, execute malicious code, or gain unauthorized access to sensitive data. Buffer overflows, for example, occur when a program writes more data to a memory buffer than it can hold, potentially overwriting adjacent memory and altering program behavior. Similarly, improper pointer handling, such as dereferencing null or uninitialized pointers, can lead to crashes or unpredictable execution.

These vulnerabilities are especially common in low-level programming languages such as C because developers are required to manually allocate, deallocate, and manage memory. Poor memory management, such as improper bounds checking or failure to free allocated memory, increases the risk of introducing security flaws. While coding best practices such as input validation and secure memory management techniques help mitigate these risks, they do not eliminate them entirely \cite{charoenwetEffectiveSecureCode2024}. Additionally, third-party dependencies can introduce vulnerabilities if they do not follow the same security standards as the rest of the codebase.

\subsection{Binary Analysis and Reverse Engineering}

One effective method to detect and prevent memory corruption vulnerabilities is through static analysis techniques, which involves analyzing a program's source code without executing it \cite{pistoiaSurveyStaticAnalysis2007}. The same methodology can be applied to compiled binaries to understand their behavior, structure, and identify potential vulnerabilities. Binary analysis is essential in situations where the source code is unavailable. For example, software vendors may not share their source code to protect intellectual property, leaving only the binary to analyze. Similarly, when analyzing malware, the source code is rarely available, requiring reverse engineers to investigate the binary to understand its functionality and structure. 

An important aspect of binary analysis is understanding a program's control and data flow. Control Flow Graphs (CFGs) are commonly used to represent possible execution paths within a program, and help visualize its behavior. In a CFG, nodes represent blocks of instructions that execute sequentially, while edges indicate possible branches in the program’s execution. 

Similarly, Data Flow Graphs (DFGs) track the movement of data through a program. This is especially useful for detecting vulnerabilities such as uninitialized variables and improper memory access; however, it is often more challenging to obtain an accurate representation than program control flow.

While these techniques are often used when source code is unavailable, binary analysis has value even when the source code can be directly analyzed. This is because vulnerabilities can be introduced during the compilation or linking stages, with one example being the dead-store elimination optimization \cite{yangDeadStoreElimination2017}.

\subsection{Deep Learning in Binary Analysis}

Several reverse engineering applications are available that provide state-of-the-art tools based on heuristics and pattern-matching \cite{ghidra, binaryninja}, however, even with these aids, the process of binary analysis remains highly time-consuming. Deep learning models have shown promise in extracting features from raw binaries, improving accuracy in tasks such as vulnerability detection and function similarity analysis \cite{arakelyanBin2vecLearningRepresentations2021, dingAsm2VecBoostingStatic2019}. 

Bin2Vec \cite{arakelyanBin2vecLearningRepresentations2021}, is an approach that learns high-dimensional embedding representations of binary programs by using Graph Convolutional Networks (GCNs), a type of neural network that is well-suited for representing the control flow relationships within binary programs.

The models were trained and evaluated using the Juliet Test Suite which contains a set of synthetic C and C++ programs with known vulnerabilities labeled according to the CWE framework \cite{CWE, bolandJuliet11Java2012}. Each test case is designed to demonstrate a specific vulnerability, including buffer overflows, use-after-free, integer overflows, and string formatting vulnerabilities. The dataset provides both vulnerable and safe code variants, allowing Bin2Vec to learn patterns from both classes.

Using the Juliet Test Suite, Bin2Vec was evaluated on two different tasks: functional algorithm classification and vulnerability discovery. For algorithm classification, Bin2Vec aims to classify programs based on functionality rather than syntactic characteristics and was shown to outperform earlier methods. For vulnerability detection, Bin2Vec achieved accuracy between 80\% and 90\% across thirty CWE classes.

Asm2Vec is another novel approach to learning embeddings from assembly language \cite{dingAsm2VecBoostingStatic2019}. Asm2Vec is designed to improve static representations for binary clone search, addressing the challenges introduced by code obfuscation and compiler optimization. 

Asm2Vec operates by using a representation learning model inspired by Word2Vec \cite{mikolovEfficientEstimationWord2013}, and represents an assembly function as a sequence of tokens, capturing both operation and operand information. By modeling the control flow graph as multiple sequences, the model learns semantic relationships between tokens through a sliding window approach. This allows the model to construct vector representations of functions based on functionality rather than syntax, even when obfuscation or optimization techniques alter the structure.
 
\section{Methodology}

\subsection{Tokenization}

Many approaches to deep learning–based binary analysis involve tokenizing assembly language or an intermediate representation, a common technique in natural language processing. (NLP) \cite{arakelyanBin2vecLearningRepresentations2021, dingAsm2VecBoostingStatic2019, liPalmTreeLearningAssembly2021} This paper takes a unique approach by directly tokenizing x86-64 machine code, which lacks the human-readable symbolic representation present in assembly language.

To tokenize machine code instructions, one strategy could involve building a vocabulary from all unique instructions from a dataset of x86-64 binaries. However, this approach would result in an enormous vocabulary size due to the wide range of possible operand bytes, causing a combinatorial explosion of unique tokens, and impacting the model's ability to learn effective embeddings.

To mitigate this issue, the operand bytes (displacement and immediate) are omitted, leaving the prefix, opcode, ModR/M, and SIB bytes \cite{Intel1997}. This approach significantly reduces vocabulary size while still preserving important information. Each of the fields from the instruction format are described in Table~\ref{table:instruction_layout}

\begin{table}[!t]
\caption{Intel x86-64 Instruction Field Layout \cite{Intel1997}}
\centering
\renewcommand{\arraystretch}{1.4}
\begin{tabularx}{\linewidth}{l c X}
\toprule
\textbf{Field} & \textbf{Size} & \textbf{Description} \\
\hline
Prefix       & 0--4 bytes       & Optional bytes placed before the opcode to modify default behavior (e.g., segment override, atomic execution, branch prediction hints). \\
\hline
Opcode       & 1--3 bytes       & One or more bytes that specify the operation to be executed (e.g., \texttt{ADD}, \texttt{MOV}, \texttt{JMP}). \\
\hline
ModR/M       & 1 byte           & A single byte encoding both the register operand and the addressing mode. \\
\hline
SIB          & 0--1 bytes       & A single byte (Scale-Index-Base) used for complex memory addressing by specifying scale factors, index registers, and base registers. \\
\hline
Displacement & 1, 2, or 4 bytes & Specifies a signed offset for memory addressing. \\
\hline
Immediate    & 1, 2, or 4 bytes & A constant value embedded in the instruction. \\
\bottomrule
\end{tabularx}
\label{table:instruction_layout}
\end{table}

The prefix field introduces details not typically seen in assembly language that could contribute to vulnerability discovery. For example, operand-size overrides could truncate values used for arithmetic operations, potentially leading to overflows. The ModR/M and SIB bytes also introduce relevant data flow information by specifying source and destination registers as well as addressing modes, even if the exact operand bytes are omitted.

\subsection{Representation}

As opposed to natural language, code is not always executed sequentially. To accommodate this, tokenized instructions will be structured into two distinct representations designed for two separate model architectures: one that simply represents each function as a single sequence, and another that represents each function as a CFG. 

\subsubsection{Sequential Representation}

The first and more straightforward of the two representations simply structures each function as a sequence of instruction tokens. This is analogous to sentences within a document in NLP, where each instruction corresponds to a word. While this approach is simple and less computationally expensive, it almost completely ignores any control flow dependencies within a function. 

Due to the variety in function and program length, functions need to be padded or truncated to a fixed length of instructions, and similarly, programs need to be padded or truncated to a fixed length of functions. The resulting structure can be represented as a matrix, with each column representing the sequence of instruction tokens for a given function. This program structure is defined more precisely below, where \(n\) is the number of instructions and \(m\) is the number of functions.	

\[
\mP_{\text{seq}} =
\begin{array}{c}  
    \begin{array}{*{6}{c}}
           (\vf_0) &  (\vf_1)  & \cdots & (\vf_{m-2}) & (\vf_{m-1})
    \end{array}
    \\[0.5ex]
    \left[
    \begin{array}{*{6}{c}}
        in_0 & in_0 & \cdots & in_0 & in_0 \\
        in_1 & in_1 & \cdots & in_1 & in_1 \\
        \vdots & \vdots & \ddots & \vdots & \vdots \\
        in_{n- 2} & in_{n - 2} & \cdots & in_{n - 2} & in_{n - 2} \\
        in_{n -1} & in_{n-1} & \cdots & in_{n-1} & in_{n-1}
    \end{array}  
    \right]
\end{array}
\]

\subsubsection{Graph-based Representation}

The sequential representation has the advantage of simplicity but does not represent control flow dependencies as effectively. The graph-based representation addresses this by constructing a CFG for each function instead of a single sequence of instruction tokens. This graph is made up of a feature matrix containing a sequence of instructions for each basic block in the function's CFG, and an adjacency matrix, representing control flow relationships between blocks (i.e. jump instructions). It is worth noting that not all control flow can be represented, as return addresses and jump operands can be calculated dynamically during runtime. This is one limitation of static analysis. 

Similar to the sequential representation, basic blocks, functions, and programs also need to be padded or truncated to a fixed size. By restricting the number of instructions per basic block and basic blocks per function to be the same, the resulting structure can be efficiently represented in a single tensor, containing pairs of feature and adjacency matrices for each function. This tensor format is defined below, where \(n\) is the number of instructions per basic block, \(m\) is the number of basic blocks per function (\(n=m\)), and \(p\) is the number of functions per program.

\[
\mF_{\text{i}} =
\begin{array}{c}  
    \begin{array}{*{6}{c}}
           (\vb_0) &  (\vb_1)  & \cdots & (\vb_{m-2}) & (\vb_{m-1})
    \end{array}
    \\[0.5ex]
    \left[
    \begin{array}{*{6}{c}}
        in_0 & in_0 & \cdots & in_0 & in_0 \\
        in_1 & in_1 & \cdots & in_1 & in_1 \\
        \vdots & \vdots & \ddots & \vdots & \vdots \\
        in_{n- 2} & in_{n - 2} & \cdots & in_{n - 2} & in_{n - 2} \\
        in_{n -1} & in_{n-1} & \cdots & in_{n-1} & in_{n-1}
    \end{array} 
    \right]
\end{array}
\]

\[\mA_i =
\begin{bmatrix}
    0 & 0 & \cdots & 1 & 0 \\
    0 & 1 & \cdots & 0 & 1 \\
    \vdots & \vdots & \ddots & \vdots & \vdots\\
    1 & 0 & \cdots & 0 & 0 \\
    0 & 1 & \cdots & 0 & 1
\end{bmatrix}
\]

\[
\mathcal{\tP}_{graph} =
\begin{bmatrix}
    \mF_0 & \mF_1 & \cdots & \mF_{p-2} & \mF_{p-1} \\
    \mA_0 & \mA_1 & \cdots & \mA_{p-2} & \mA_{p-1}
\end{bmatrix}
\]

\subsection{Models}

For each representation discussed, a supervised binary classification model was developed. Each model constructs hierarchical embeddings with progressively decreasing granularity. Both models learn static instruction embeddings, which provide a mapping from each instruction token to a continuous vector space. The sequential model then extracts features from the instruction embeddings to create function-level embeddings, extracts features from the function-level embeddings to create a program-level embedding, and finally makes a prediction based on the program-level embedding. The graph-based model has a similar design except for the additional intermediate stage of basic block embeddings within each function's CFG. 

\subsubsection{Sequential Architecture}

The sequential model architecture is made up of four main components: An instruction embedding layer, a one-dimensional convolutional neural network (1D-CNN) layer for extracting features from adjacent instructions, a self-attention layer for aggregating functions, and a fully connected layer to make final classifications. 

The instruction embedding layer first maps each token \(in_i\) to a learned vector \(\vx_{in_i}\), which is randomly initialized. To extract local patterns from adjacent instructions, a single 1D-CNN layer was used. 1D-CNNs have shown potential in sentence classification, motivating the decision to apply it to instruction sequences \cite{kimConvolutionalNeuralNetworks2014}. This generates a function-level embedding, \( \vx_{func_j}\), containing the extracted and aggregated features from the instruction sequence.

Under the assumption that most features indicative of a vulnerability exist in a small number of functions, a self-attention layer is used to attend to and then aggregate the functions \cite{vaswaniAttentionAllYou2023a}. This encourages the model to focus on functions that contribute most to the final classification.

Given a set of function embeddings, \(\mX_{func}\), a multi-head attention layer is used to obtain a score matrix, \(\mS\), representing how much each function attends to every other function. \(\mS\) is used to compute a weighted representation of \(\mX_{func}\) that has been updated with contextual information from the other functions. An additional advantage of this approach is that the attention scores can potentially provide transparency into which function(s) contribute most to a given vulnerability.

The attended function embeddings are then aggregated using mean pooling, and projected to a single program embedding, \(\vx_{prog} \in \mathbb{R}^{prog\_dim}\)

Finally, a feedforward network performs the final classification on the final program embedding. This network contains an input layer of size \(prog\_dim\) with decreasing hidden layer sizes and an output layer with a single neuron. A sigmoid activation is then applied to convert the logit to a probability.

\subsubsection{Graph-based Architecture}

The graph-based architecture has a similar design, except with an intermediate graph convolutional network (GCN) for handling each function's control flow graph \cite{kipfSemiSupervisedClassificationGraph2017a}. In this model, the input tensor, \( \mX \), contains pairs of feature and adjacency matrices, where \( \mX[:,0,i, :, :] \) is a feature matrix for function \( i \), containing column vectors for each basic block, and \( \mX[:,1,i, :, :] \) represents the corresponding adjacency matrix for the function's CFG.

The graph-based architecture first learns static instruction embeddings using the same method as the sequential model. A 1D-CNN is then used to extract features from each basic block’s instruction sequence, rather than entire functions. 

After obtaining an embedding for each basic block within each function, a GCN is then used to propagate information from neighboring basic blocks to learn representations that capture control flow dependencies within a function. Each GCN layer, \( l \), is defined as follows:

\[
\mH_i^{l+1} = ReLU \left( \tilde{\mD}_i^{-\frac{1}{2}} \tilde{\mA}_i \tilde{\mD}_i^{-\frac{1}{2}} \mH_i^l \mW^l \right)
\]

Where \( \mH_i^l \) is the \( i^{th} \) function’s feature matrix at layer \( l \), with \( \mH^0 = \mX[:,0,i, :, :] \). The matrix \( \tilde{\mA} = \mA + \mI \) is the corresponding adjacency matrix with added self-loops, where \( \mA = \mX[:,1,i, :, :] \). The diagonal degree matrix \( \tilde{\mD} \) is defined as \( \tilde{\mD}_{ii} = \sum_j \tilde{\mA}_{ij} \), and \( \mW^l \) is a layer-specific trainable weight matrix. Here, the term \( \tilde{\mD}_i^{-\frac{1}{2}} \tilde{\mA}_i \tilde{\mD}_i^{-\frac{1}{2}} \) represents the symmetric normalization of the adjacency matrix, which is used to improve stability and prevent feature explosion \cite{kipfSemiSupervisedClassificationGraph2017a}. 

To create a function-level embedding, each updated block in the CFG must be aggregated or pooled. This is a challenging task as most functions produce sparse CFGs, and many implement common routines containing similar control flow patterns. To address this, an attention-based approach inspired by \cite{cangeaSparseHierarchicalGraph2018} was used. This approach involves only pooling the top \(K\) nodes in the graph, based on an attention score assigned to each node.

The attention scores are computed based on a simple feedforward network applied to each node's features, with a single hidden layer, and a single output neuron. A softmax function is then applied across each node's score, \(\vs_i\), with an additional temperature parameter, \(t\) \cite{xuan2025exploringimpacttemperaturescaling}.

\[
\boldsymbol{\alpha_i} = \frac{\exp(\vs_i / t)}{\sum_{j}^{num\_nodes} \exp(\vs_j / t)}
\]

Using the attention scores of each node, \( \boldsymbol{\alpha_i}\), the indices of the top \( K \) nodes are selected and extracted from the resulting feature matrix from the final GCN layer:

\[
I = \arg \text{topK}(\boldsymbol{\alpha}), \quad \mH_{topK} = \mathbf{H}^{final}[:, :, I, :]
\]

Here, the columns of \( \mH_{topK}[:, i, :, :] \) represent the most important CFG blocks of function \( i \). Each of these block embeddings is then concatenated and projected to a function-level embedding.

The same self-attention layer and final feedforward layer described in the sequential architecture are used to aggregate these function embeddings into a single program-level embedding and perform the final classification.

\section{Experiment}

This paper focuses on detecting three specific vulnerability types: Null Pointer Dereference errors (CWE-476), Improper Validation of Array Index (CWE-129), and Integer Overflow or Wraparound (CWE-190). These vulnerabilities were selected to represent three broader classes of software flaws: pointer errors, buffer errors, and numeric errors. Designing classifiers for each can offer insight into detecting related vulnerabilities in future research. 

\subsection{Dataset}

Bin2Vec demonstrated promising results in vulnerability detection by training on the Juliet Test Suite \cite{bolandJuliet11Java2012}, a dataset composed of synthetic code examples to isolate specific vulnerabilities. While the Juliet Suite remains a valuable resource for testing static analysis tools, it may not generalize well in the context of machine learning. In contrast, the FormAI-v2 dataset \cite{tihanyiHowSecureAIgenerated2025a} aims to include more realistic code by generating C programs using Large Language Models (LLMs) without artificially restricting the presence or number of vulnerabilities. The resulting code is then formally verified to identify genuine security issues, including scenarios where multiple vulnerabilities coexist.

The dataset was created by prompting nine different LLMs to generate code containing a variety of tasks and coding styles, ranging from simple string manipulation to network analysis and encryption. Each of the resulting 331,000 programs were then tested for compilability using the GNU C compiler to ensure that any code in the dataset could be compiled in a standard environment.

To label and categorize vulnerabilities accurately, the dataset utilized the Efficient SMT-based Bounded Model Checker (ESBMC) \cite{gadelhaESBMC50Industrialstrength2018}, which examines possible execution paths and flags potential security issues. Overall, approximately 62\% of the generated programs were labeled as vulnerable, reflecting the reality that LLM-generated code often contains security flaws. After compilation checks and verification, any clones present in the dataset were removed using the NiCad clone detection tool. \cite{cordyNiCadCloneDetector2011}.

\subsection{Preprocessing}

Due to the nature of the violation types included in the FormAI-v2 dataset, the distribution is highly imbalanced. To address this, similar types of violations were grouped in the same vulnerability class. Specifically, upper bound and lower bound variants of array bound violations were generalized into a single class, and similarly, arithmetic overflows on SUB, ADD and MUL were also combined. Due to the high number of null pointer dereference failures, this was represented as its own class.

Three separate subsets were extracted from the complete dataset for each vulnerability class. A binary was labeled as vulnerable if that specific type of violation was present, otherwise, it was labeled as non-vulnerable, regardless of whether another type of violation was present. 

Additionally, to ensure a balance within each vulnerability class, a 50/50 split between vulnerable and non-vulnerable files was selected. Each subset was then split into an 80/10/10 train/validation/test split. The resulting dataset of source code files is outlined in Table~\ref{table:source_code_dist}

\begin{table}[!t]
    \centering
    \renewcommand{\arraystretch}{1.4}
    \caption{Source Code Distribution by Vulnerability Class}
    \label{table:source_code_dist}
    \begin{tabularx}{\linewidth}{l c c c c}
        \toprule
        \textbf{Vulnerability Class} & \textbf{CWE ID} & \textbf{Count} & \textbf{Set} & \textbf{Count} \\
        \hline
        Null Pointer Dereference & CWE-476 & 100,436 & Train & 80,348 \\
                                 &         &         & Val   & 10,046 \\
                                 &         &         & Test  & 10,042 \\
        \hline
        Array Bound Violation    & CWE-129 & 25,068  & Train & 20,054 \\
                                 &         &         & Val   & 2,508 \\
                                 &         &         & Test  & 2,506 \\
        \hline
        Integer Overflow         & CWE-190 & 39,996  & Train & 31,996 \\
                                 &         &         & Val   & 4,002 \\
                                 &         &         & Test  & 3,998 \\
        \bottomrule
    \end{tabularx}
\end{table}

To address the remaining imbalance, each file was compiled multiple times with different compiler optimizations, proportional to the degree of imbalance. Specifically, the GNU C compiler was used, and at each instance a file was compiled, an optimization flag was selected at random without replacement to ensure the same file was not repeatedly compiled with the same flag. The optimization flags used include -O0, -O1, -O2, -O3, -Os, and -Ofast, each providing varying degrees of optimization in terms of execution time, code size, memory usage, and compile time \cite{GCCOptimizeOptions}. It is not guaranteed that the resulting binaries from the same source code will have significant differences; hence, the train/validation/test sets were created before compilation, to avoid data leakage and skewing evaluation results

This approach improves the balance between vulnerability classes, augments the dataset size significantly, and introduces more diverse machine code, which could potentially improve predictive performance on optimized and obfuscated code. After the compilation process, each dataset was downsampled to the minimum size to ensure a fair comparison between vulnerability classes. The final dataset is presented in Table~\ref{table:final_dist}.

\begin{table}[!t]
\centering
\footnotesize                % reduce font size to fit in one column
\renewcommand{\arraystretch}{1.2}
\setlength{\tabcolsep}{3pt}  % narrow column spacing
\caption{Resulting Dataset Distribution After Compilation}
\label{table:final_dist}

\begin{tabularx}{\linewidth}{l c c c c}
\toprule
\makecell{\textbf{Class}\\[3pt]} & 
\makecell{\textbf{Before}\\ \textbf{Compilation}} & 
\makecell{\textbf{Times}\\ \textbf{Compiled}} & 
\makecell{\textbf{After}\\ \textbf{Compilation}} & 
\makecell{\textbf{After}\\ \textbf{Downsampling}} \\
\midrule
\makecell[l]{Null Pointer\\Dereference} & 
100,436 & $\times2$ & 200,872 & 150,408 \\
\hline
\makecell[l]{Array Bound\\Violation} & 
25,068  & $\times6$ & 150,408 & 150,408 \\
\hline
\makecell[l]{Arithmetic\\Overflow} & 
39,996  & $\times4$ & 159,984 & 150,408 \\
\bottomrule
\end{tabularx}
\end{table}

\subsection{Model Hyperparameters}

Most hyperparameters were tuned manually on small validation subsets. However, the CNN filter size in the sequential models and the GCN layer depth in the graph-based models were subject to a more systematic hyperparameter search over the full dataset, as detailed in the following sections.

\subsubsection{Sequential Model Hyperparameters}

To gain further insight into the sequential architecture, four variants were trained and evaluated. Three models used single filter sizes of 3, 5, and 7 (each with 768 filters), and a hybrid model combined all three sizes with 256 filters per branch. Each convolutional branch is followed by batch normalization, ReLU activation, 30\% dropout, and global average pooling.

The instruction- and function-level embedding dimensions were set to 512, while the program-level embedding dimension was 256. The multi-head attention layer that aggregates function embeddings was configured with eight heads, and the final feed-forward network contained three hidden layers of sizes 256, 128, and 64.

\subsubsection{Graph-based Model Hyperparameters}

In the graph-based architecture, the 1D-CNN layer used 512 filters of size 3 because it operates only within individual basic blocks rather than across entire functions, and thus required fewer or smaller filters to capture meaningful patterns. Similar to the sequential model's CNN, this is followed by batch normalization, ReLU activation, 30 \% dropout, and global average pooling.

Because the distinguishing component of this architecture is the GCN that models relationships between basic blocks, four variants with one to four GCN layers were trained to assess the effect of depth. A hidden dimension of 512 is used between layers. After the final GCN layer, top-\(K\) pooling (\(K=0.1\), \(t=0.1\)) produces function-level embeddings.

The eight-head attention layer for aggregating functions and the final feed-forward network are identical to those in the sequential model. Instruction-, basic-block-, and function-level embedding dimensions were set to 512, and the program-level embedding dimension was set to 256.

\subsection{Training Hyperparameters}

All models were trained using the same training hyperparameters to ensure a fair comparison. The Adam optimization algorithm was used with a learning rate of \( 5 \times 10^{-5} \) and weight decay of \( 1 \times 10^{-5} \). Binary cross-entropy was used as the loss function, and to stabilize training, gradient clipping was also applied.

The batch size was set to 8 for graph-based models and 32 for sequential models, chosen based on available computational resources to balance memory efficiency and training stability. Early stopping was applied with \( \text{patience} = 3 \) and \( \delta = 0 \) to halt training when no further improvements were observed. A threshold of 0.5 was used to classify model outputs, where an output \( \geq 0.5 \) indicates a vulnerable sample, and an output \( < 0.5 \) indicates a non-vulnerable sample.

A holdout validation approach was used on the training, validation, and test sets, where the validation set was used to monitor training performance, tune hyperparameters, and detect overfitting. The test set remained unseen by the models during training and was used exclusively for final model evaluation to ensure a fair assessment of generalization.

\section{Results}

To evaluate the impact of convolutional filter size on performance within the sequential architecture, experiments were conducted with different filter sizes, \( k = 3 \), \( k = 5 \), \( k = 7 \), and a combination of filter sizes, \( k \in \{3,5,7\} \). The results for accuracy and F1-score across the three vulnerability types are illustrated in Table~\ref{tab:filter_sizes}. Increasing the filter size generally improved performance, with \(k=7\) resulting in the highest accuracy and F1-score across all vulnerability types.

\begin{table}[ht]
\caption{Effects of Filter Size on Model Performance}
\centering
\renewcommand{\arraystretch}{1.2}
\begin{tabularx}{\linewidth}{c c >{\centering\arraybackslash}X >{\centering\arraybackslash}X >{\centering\arraybackslash}X}
\hline
\textbf{\textit{k}} & \textbf{Metric} & \textbf{Null Pointer Dereference} & \textbf{Array Bound Violation} & \textbf{Integer Overflow} \\
\hline
3 & Accuracy & 0.7683 & 0.7155 & 0.6969 \\
  & F1-Score & 0.7558 & 0.7336 & 0.6932 \\
\hline
5 & Accuracy & 0.8001 & 0.7399 & 0.7166 \\
  & F1-Score & 0.7993 & 0.7503 & 0.7284 \\
\hline
7 & Accuracy & \textbf{0.8152} & \textbf{0.7584} & \textbf{0.7334} \\
  & F1-Score & \textbf{0.8145} & \textbf{0.7617} & \textbf{0.7339} \\
\hline
\{3,5,7\} & Accuracy & 0.7827 & 0.7336 & 0.7053 \\
  & F1-Score & 0.7803 & 0.7467 & 0.7049 \\
\hline
\end{tabularx}
\label{tab:filter_sizes}
\end{table}

The effect of varying the number of Graph Convolutional Network (GCN) layers on classification performance was also examined. As shown in Table~\ref{tab:gcn_layers}, performance generally declined as additional layers were introduced beyond a certain point, with three layers resulting in the worst performance. A two-layer GCN achieved the best performance in classifying null pointer dereference errors and array bound violations, while the one-layer GCN configuration achieved the best performance in classifying integer overflows.

\begin{table}[ht]
\caption{Effects of GCN Layers on Model Performnace}
\centering
\renewcommand{\arraystretch}{1.2}
\begin{tabularx}{\linewidth}{c c >{\centering\arraybackslash}X >{\centering\arraybackslash}X >{\centering\arraybackslash}X}
\hline
\textbf{Layers} & \textbf{Metric} & \textbf{Null Pointer Dereference} & \textbf{Array Bound Violation} & \textbf{Integer Overflow} \\
\hline
1 & Accuracy & 0.8998 & 0.8568 & \textbf{0.8392} \\
  & F1-Score & 0.9024 & 0.8612 & \textbf{0.8389} \\
\hline
2 & Accuracy & \textbf{0.9323} & \textbf{0.8661} & 0.8386 \\
  & F1-Score & \textbf{0.9332} & \textbf{0.8654} & 0.8388 \\
\hline
3 & Accuracy & 0.7647 & 0.8256 & 0.7811 \\
  & F1-Score & 0.7770 & 0.8121 & 0.7843 \\
\hline
4 & Accuracy & 0.8204 & 0.8632 & 0.8204 \\
  & F1-Score & 0.8110 & 0.8614 & 0.8110 \\
\hline
\end{tabularx}
\label{tab:gcn_layers}
\end{table}

\begin{figure}[h] 
    \centering
    \includegraphics[width=1.0\linewidth]{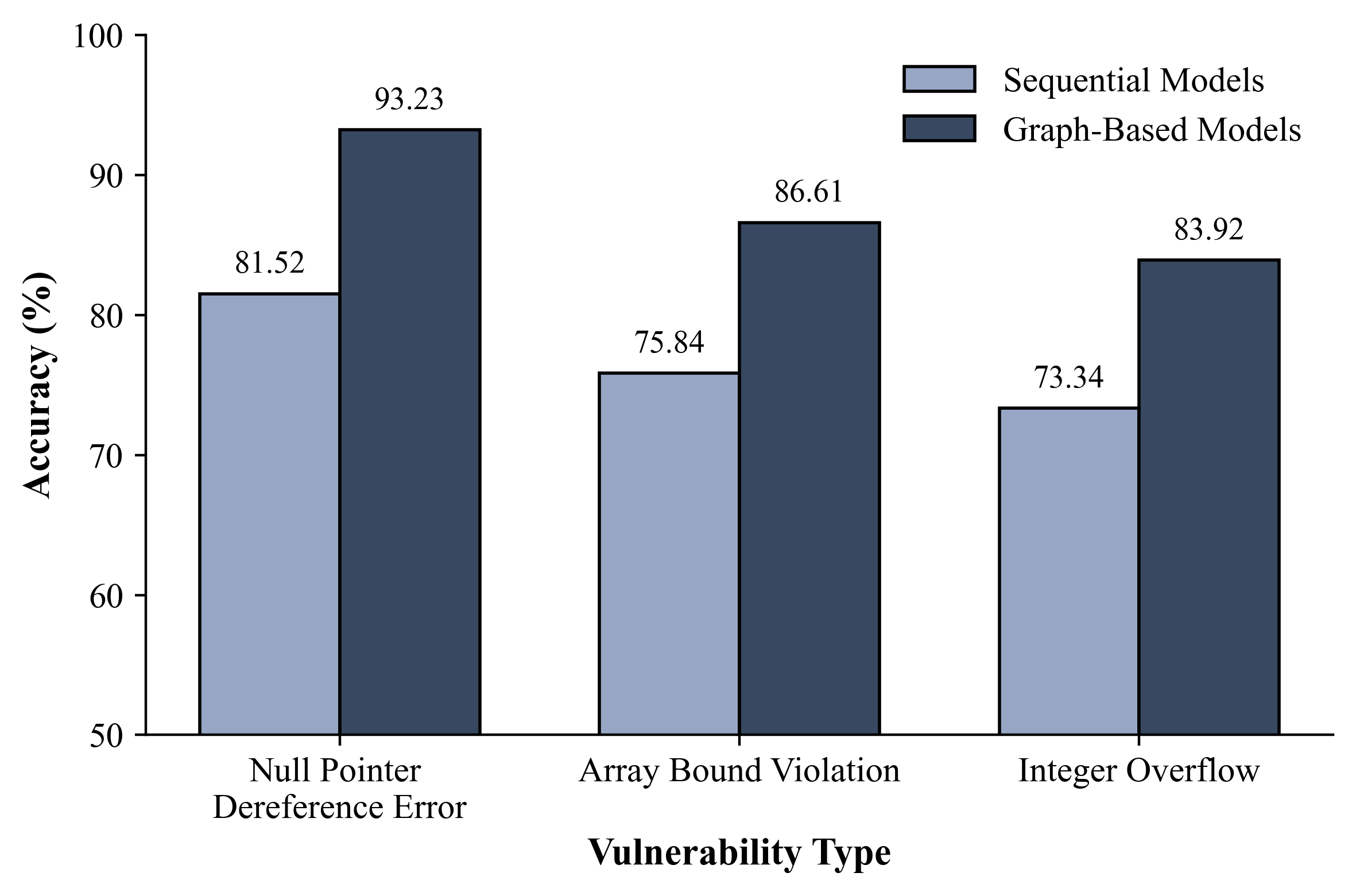}
    \caption{Top sequential and graph-based models across vulnerability types}
    \label{figure:top_models}
\end{figure}

Lastly, the top-performing sequential models are compared against the top-performing graph-based models. For the sequential models, the best-performing models were those configured with a kernel size of 7 in the CNN layer across all vulnerability classes. For the graph-based models, the two-layer GCN performed best in classifying null pointer dereference errors and array bound violations, while the single-layer GCN performed best in classifying Integer Overflows. These top models are presented in Figure~\ref{figure:top_models}.

It is also worth noting that across all model configurations, null pointer dereference error classifiers generally achieved the highest performance, while integer overflow classifiers performed the worst.

\section{Discussion}

One of the most notable observations from the results is the consistent outperformance of graph-based models over sequential models across all vulnerability types. This indicates that the adjacency relationships between instructions are not enough to fully capture the dependencies leading to security flaws. The improved performance of graph-based models demonstrates the importance of control flow relationships rather than relying only on local instruction sequences. 

Despite the effectiveness of graph convolutional network (GCN) models, and their ability to incorporate control flow dependencies, the results indicate that adding too many GCN layers can degrade performance. This could suggest the occurrence of over-smoothing, a phenomenon in deep graph networks where excessive message passing causes node embeddings to become indistinguishable \cite{congProvableBenefitsDepth2021}. 

Another insight is that increasing convolutional filter sizes generally improves performance, but combining multiple filter sizes did not lead to noticeable improvements. This could indicate that an optimal filter size exists for different types of vulnerabilities. The lack of significant improvement when mixing multiple filter sizes further supports the idea that configuring a model to specific vulnerability types is more important than aiming to address a broad range of vulnerabilities with a single model through excessive parameterization. 

Another observation is that integer overflow classification consistently showed the lowest performance across all models. Unlike a null pointer dereference error or array bound violation, integer overflows often do not produce explicit control flow anomalies. Instead, they depend on specific arithmetic operations that exceed fixed-width integer limits, which may be difficult to detect from control flow graphs alone. Traditional static analysis tools often struggle with integer overflows for similar reasons, as they require symbolic execution or taint analysis to track numerical constraints across program execution paths \cite{schwartzAllYouEver2010}.

The results indicate that the models achieve similar performance as Bin2Vec, which reports 80-90\% accuracy across multiple CWEs. This performance is achieved without utilizing rich semantic information from assembly language or an intermediate representation, relying solely on x86 machine code instruction encodings. 

\subsection{Limitations and Future Work}

While this study demonstrates the feasibility of using deep learning for vulnerability discovery in machine code, there are several directions for future research to improve and extend this work.

Firstly, this study is only limited to three vulnerability types; however, software can contain a much broader range of issues, such as race conditions and use-after-free errors. Future research could extend the model’s capabilities to detect additional vulnerability types.

Another limitation is the lack of dataset diversity in model training and evaluation. While FormAI provides a diverse set of C code spanning various styles and domains, it is restricted to Linux-based C libraries. To develop a more robust model, training should incorporate binaries from multiple operating systems but will need to consider different binary formats such as PE files for Windows and ELF files for Linux. Incorporating binaries from different operating systems will also need to consider inconsistencies in system calls, as identical instructions may correspond to entirely different operations depending on each operating system's system call convention.

The models presented in this paper rely exclusively on local instruction relationships and control flow information to represent binaries, but vulnerabilities often result from unsafe data flow rather than control flow patterns. This limitation is supported by the poor performance of integer overflow classification, emphasizing the importance of data flow dependencies. 

Another limitation of this approach is the lack of interpretability. In real-world applications, security professionals must trust and understand model decisions to mitigate risks effectively. Therefore, future work could explore techniques to improve interpretability and potentially locate insecure code within a binary, greatly improving the tedious task of reverse engineering. The use of self-attention in the models presented in this paper can provide the foundation for achieving this, as it provides attention scores for specific basic blocks or functions, helping locate vulnerable regions. 

Lastly, by learning representations directly from machine code, the scope is narrowed to a single architecture. Future research could investigate applying this approach to more architectures, such as ARM or Java Bytecode. This could reveal specific architectures that benefit more from direct machine code analysis. For example, reduced instruction sets (RISC) could benefit from this approach due to a smaller instruction vocabulary.

\section{Conclusion}

This paper investigated the effectiveness of deep learning-based vulnerability detection methods using raw x86-64 machine code, addressing a gap in existing research primarily focusing on assembly language or intermediate representations. 

An important contribution of this work is the finding that meaningful vulnerability patterns can indeed be extracted directly from machine code and instruction encodings without relying on richer semantic representations. The models presented achieved accuracy comparable to existing methods reliant on assembly language or an intermediate representation, demonstrating that machine code alone provides sufficient information for effective vulnerability detection. This result suggests promising directions for developing simpler, less resource-intensive models that maintain high accuracy levels, applicable for real-world scenarios where computational resources are limited.

Future research directions include expanding the range of vulnerability types, incorporating data-flow analysis, and applying this approach to different system architectures. Additionally, enhancing model interpretability would improve real-world security analysis and decision-making.

Overall, this research demonstrates the feasibility of using machine code representations for vulnerability detection. It provides a foundation for future advancements in efficient, interpretable, and robust deep learning approaches for binary analysis.

\section*{Acknowledgment}

I would like to thank Dr. George Townsend and Dr. Randy Lin for their thorough reviews and valuable comments. Their feedback greatly improved the quality of this work.

\bibliographystyle{IEEEtran}   
\bibliography{refs}   

\end{document}